\documentclass[a4paper,11pt]{article}
\usepackage[centertags]{amsmath}
\usepackage{amssymb,amsmath,amsthm,amscd,latexsym}
\usepackage{amsfonts}
\usepackage{tikz}
\usetikzlibrary{arrows,positioning,shapes,fit,calc}
\usepackage{tikz-cd}
\usepackage[mathscr]{eucal}
\usepackage{color,hyperref}
\usepackage{graphicx}
\usepackage{multirow}
\usepackage[ruled,vlined]{algorithm2e}
\usepackage{xcolor}
\usepackage{amsfonts}
\usepackage{amssymb}
\usepackage{amsthm}

 \theoremstyle{definition}
 
 \theoremstyle{remark}

 \numberwithin{equation}{section}
\begin{document}

\title{An Application of the Virus Optimization Algorithm to the Problem of Finding Extremal Binary Self-Dual Codes}
\author{Adrian Korban \\
Department of Mathematical and Physical Sciences \\
University of Chester\\
Thornton Science Park, Pool Ln, Chester CH2 4NU, England \\
Serap \c{S}ahinkaya \\
Tarsus University, Faculty of Engineering \\ Department of Natural and Mathematical Sciences \\
Mersin, Turkey \\
Deniz Ustun \\
Tarsus University, Faculty of Engineering \\ Department of Computer Engineering \\
Mersin, Turkey}
 \maketitle

\begin{abstract}

In this paper, a virus optimization algorithm, which is one of the metaheuristic optimization technique, is employed for the first time to the problem of finding extremal binary self-dual codes. We present a number of generator matrices of the form $[I_{36} \ | \ \tau_3(v)],$ where $I_{36}$ is the $36 \times 36$ identity matrix, $v$ is an element in the group matrix ring $M_3(\mathbb{F}_2)G$  and $G$ is a finite group of order 12, which we then employ together with the the virus optimization algorithm and the genetic algorithm to search for extremal binary self-dual codes of length 72. We obtain that the virus optimization algorithm finds more extremal binary self-dual codes  than the genetic algorithm. Moreover, by employing the above mentioned constructions together with the virus optimization algorithm, we are able to obtain 39 Type I and 19 Type II codes of length 72, with parameters in their weight enumerators that were not known in the literature before.
\end{abstract}

\textbf{Key Words}:binary self-dual codes, virus optimization algorithm, metaheuristic optimization, extremal codes

\section{Introduction}

Self-dual codes over finite fields, especially binary fields, have been one of the most important and widely studied subject in algebraic coding theory.
Self-dual codes over rings have been shown to have many interesting connections to invariant
theory, lattice theory and the theory of modular forms, see \cite{Gebe}. Classification of binary self-dual codes of different lengths is still an ongoing research area in algebraic coding theory.
In this context, various techniques have been employed to search for binary self-dual codes of different lengths with new parameters in their weight enumerators. A generator matrix of the form $[I_n \ | \ A_n],$ where $I_n$ is the $n \times n$ identity matrix and $A_n$ is some $n \times n$ matrix with entries from a finite field $\mathbb{F}_2$ is the most common technique for producing extremal binary self-dual codes. Later, this technique has been extended in  \cite{Gildea1}, by replacing the matrix $A$ with $\sigma(v)$, where  $\sigma(v)$ is the image of a unitary unit in a group ring under a map that sends group ring elements to matrices that are fully defined by the elements appearing in the first row. One can see \cite{Dougherty3,Dougherty4,Dougherty5} for more applications of this technique. In \cite{CompositeGCodes}, the map $\sigma(v)$ was extended to $\Omega(v)$, which gives more complex matrices over the ring $R,$ and  generator matrices of the form $[I_n \ | \ \Omega(v)]$ were considered.
Recently in \cite{Dougherty3}, the map $\sigma$ was extended $\tau_k$ by considering elements from the group matrix ring $M_k(R)G$ rather than elements from the group ring $RG$, and
generator matrices of the form $[I_{kn} \ | \tau_k(v)]$ were considered to produce binary self-dual codes. Please see \cite{Dougherty3} for the details of this construction. The advantage of the map $\tau_k$ is that it does not only depend on the choice of the group $G,$ but it also depends on the form of the elements from the matrix ring $M_k(R),$ that is, the form of the $k \times k$ matrices over $R.$

Metaheuristic optimization algorithms have been widely used successfully  for many engineering problems in which solution steps take exhaustively long time \cite{Ustun, Karaboga3, Liang, Holland, Deniz3}.  In terms of algebraic coding theory,  optimization algorithms were used  only for  the minimum distance problem \cite{Shenoy,Bland1, Bland2, Bouzkraoui}. Searching for self-dual codes with
new parameters is one of the main problem in algebraic coding theory and linear search was the only search tool for this problem.  Although linear search for self-dual codes achieve
good results for small size search space, this is really time consuming when the search field grows. Therefore, the problem of searching for extremal self-dual codes  needs some new tools for the big size search space.
Recently, in \cite{Korban1}, the authors employed the genetic algorithm (GA), one of the well-known optimization algorithm, for the first time  to the problem of finding new binary self-dual codes. It was proved that
GA copes with large search fields significantly better and finds codes much faster then the standard linear search.

In this work,  a virus optimization algorithm (VOA) \cite{Cuevas} is employed for the first time to the problem of finding extremal binary self-dual codes of length 72.  We consider the generator matrices of the form $[I_{36} \ | \tau_3(v)]$  together with the VOA and  search for binary self-dual codes with parameters $[72,36,12].$ We find such codes with weight enumerators that were not known in the literature before. It is known by \cite{Liang} that VOA outperforms the GA in continuous problems. Therefore we  compare the number of codes found by VOA and GA for the same constructions and obtain that  VOA finds more binary self-dual codes  than GA.

The rest of the work is organized as follows. In Section~2, we give preliminary definitions and results on self-dual codes,  group rings and  recall the map $\tau_k(v)$ that was defined in \cite{Dougherty3}. In Section~3, we present a number of generator matrices of the form $[I_{36} \ | \ \tau_3(v)]$ and for each generator matrix, we fix the $3 \times 3$ matrices by letting them be some special matrices that we define in Section~2.
In Section~4, we give a brief history of VOA and explain how to employ this algorithm to the problem of finding extremal binary self-dual codes. In Section~5, we employ the generator matrices from Section~4 and search for binary self-dual codes with parameters $[72,36,12].$ As a result we find 39 Type I and 19 Type II binary $[72,36,12]$ self-dual codes with parameters in their weight enumerators that were not previously known. We tabulate our results, stating clearly the parameters of the obtained codes and their orders of the automorphism group. We finish with concluding remarks and directions for possible future research.

\section{Preliminaries}

In this section we recall some well-known definitions and  notions from algebraic coding theory.

 A code $C$ of length $n$ over a Frobenius ring $R$ is a subset of $R^n$. If the code is
a submodule of $R^n$ then we say that the code is linear. Let $\mathbf{x}=(x_1,x_2,\dots,x_n)$
and $\mathbf{y}=(y_1,y_2,\dots,y_n)$ be two elements of $R^n.$ Then
\begin{equation*}
\langle \mathbf{x},\mathbf{y} \rangle_E=\sum x_iy_i.
\end{equation*}
The dual $C^{\bot}$ of the code $C$ is defined as
\begin{equation*}
C^{\bot}=\{\mathbf{x} \in R^n \ | \ \langle \mathbf{x},\mathbf{y}
\rangle_E=0 \ \text{for all} \ \mathbf{y} \in C\}.
\end{equation*}
We say that $C$ is self-orthogonal if $C \subseteq C^\perp$ and is self-dual
if $C=C^{\bot}.$

An upper bound on the minimum Hamming distance of a binary self-dual code
was given in \cite{Rains1}. Let $d_{I}(n)$ and $d_{II}(n)$ be the
minimum distance of a Type~I (singly-even) and Type~II (doubly-even) binary code of length $n$,
respectively. Then
\begin{equation*}
d_{II}(n) \leq 4\lfloor \frac{n}{24} \rfloor+4
\end{equation*}
and
\begin{equation*}
d_{I}(n)\leq
\begin{cases}
\begin{matrix}
4\lfloor \frac{n}{24} \rfloor+4 \ \ \ if \ n \not\equiv 22 \pmod{24} \\
4\lfloor \frac{n}{24} \rfloor+6 \ \ \ if \ n \equiv 22 \pmod{24}.%
\end{matrix}%
\end{cases}%
\end{equation*}

Self-dual codes meeting these bounds are called \textsl{extremal}.

A circulant matrix is one where each row is shifted one element to the right relative to the preceding row and a reverse circulant
matrix is one where each row is shifted one element to the left relative to the preceding row.
We label the circulant matrix as $A=circ(\alpha_1,\alpha_2\dots , \alpha_n),$ and the reverse circulant matrix as $A=revcirc(\alpha_1,\alpha_2\dots , \alpha_n),$  where $\alpha_i$ are ring elements.
The transpose of a matrix $A,$ denoted by $A^T,$ is a matrix whose rows are the columns of $A,$ i.e., $A^T_{ij}=A_{ji}.$

 Let $G$ be a finite group of order $n$, then the group ring $RG$
consists of $\sum_{i=1}^n \alpha_i g_i$, $\alpha_i \in R$, $g_i \in G.$

Addition in the group ring is done by coordinate addition, namely
\begin{equation}
\sum_{i=1}^n \alpha_i g_i +\sum_{i=1}^n \beta_i g_i =\sum_{i=1}^n (\alpha_i
+ \beta_i)g_i.
\end{equation}
The product of two elements in a group ring is given by
\begin{equation}
\left(\sum_{i=1}^n \alpha_i g_i \right)\left(\sum_{j=1}^n \beta_j g_j
\right)= \sum_{i,j} \alpha_i \beta_j g_i g_j.
\end{equation}
It follows that the coefficient of $g_k$ in the product is $\sum_{g_i
g_j=g_k} \alpha_i \beta_j.$

We now recall the map $\tau_k(v),$ where $v \in M_k(R)G$ and where $M_k(R)$ is a non-commutative Frobenius matrix ring and $G$ is a finite group of order $n,$ that was introduced in \cite{Dougherty3}.

Let $v=A_{g_1}g_1+A_{g_2}g_2+\dots+A_{g_n}g_n \in M_k(R)G,$ that is, each $A_{g_i}$ is a $k \times k$ matrix with entries from the ring $R.$ Define the block matrix $\sigma_k(v) \in (M_{k}(R))_n$ to be

\begin{equation}\label{sigmakv}
\sigma_k(v)=
\begin{pmatrix}
A_{g_1^{-1}g_1} & A_{g_1^{-1}g_2} & A_{g_1^{-1}g_3} & \dots &
A_{g_1^{-1}g_{n}} \\
A_{g_2^{-1}g_1} & A_{g_2^{-1}g_2} & A_{g_2^{-1}g_3} & \dots &
A_{g_2^{-1}g_{n}} \\
\vdots & \vdots & \vdots & \vdots & \vdots \\
A_{g_{n}^{-1}g_1} & A_{g_{n}^{-1}g_2} & A_{g_{n}^{-1}g_3} & \dots &
A_{g_{n}^{-1}g_{n}}
\end{pmatrix}
.
\end{equation}

We note that the element $v$ is an element of the group matrix ring $M_k(R)G.$

{\bf Construction 1}
For a given element $v \in M_k(R)G,$ we define the following code over the matrix ring $M_k(R)$:
\begin{equation}
C_k(v)=\langle \sigma_k(v) \rangle.
\end{equation}
Here the code is generated by taking the all left linear combinations of the rows of the matrix with coefficients in $M_k(R).$

{\bf Construction 2}
For a given element $v \in M_k(R)G,$ we define the following  code over the  ring $R$.
Construct the matrix $\tau_k(v)$ by viewing each element in a $k$ by $k$ matrix as an element in the larger matrix.
\begin{equation}
B_k(v)=\langle \tau_k(v) \rangle.
\end{equation}
Here the code $B_k(v)$ is formed by taking all linear combinations of the rows of the matrix with coefficients in $R$.
In this case the ring over which the code is defined is commutative so it is both a left linear and right linear code.

 Later in this work, we employ this map and consider generator matrices of the form $[I_{kn} \ | \ \tau_k(v)]$ for groups of order $12$ and for $k=3.$ That is, we consider $3 \times 3$ matrices of different forms.

\section{Main Construction}

Let $v \in M_3(\mathbb{F}_2)G$. In this work we consider a generator matrix of the form

$$G=[I_{36} \ | \ \tau_3(v)],$$

where $G$ is a group of order $12$ and $\tau_3(v)$ is the $36 \times 36$ matrix that consists of some 12 block matrices of size $3 \times 3$. For this work, we only consider the following cases:

  \begin{equation}\label{nine}
\begin{aligned}
A_1=circ(a_1,a_2,a_3),\dots, A_{12}=circ(a_{34},a_{35},a_{36}).
\end{aligned}
\end{equation}

 \begin{equation}\label{two}
 A_1=revcirc(a_1,a_2,a_3), \dots, A_{12}=revcirc(a_{34},a_{35},a_{36}).
\begin{aligned}
\end{aligned}
\end{equation}

 \begin{equation}\label{five}
\begin{aligned}
A_1=revcirc(a_1,a_2,a_3), \dots, A_6=revcirc(a_{16},a_{17}, a_{18}), \\
A_7=circ(a_{19},a_{20}, a_{21}), \dots, A_{12}=circ(a_{34},a_{35},a_{36}).
\end{aligned}
\end{equation}

 \begin{equation}\label{one}
\begin{aligned}
A_1=circ(a_1,a_2,a_3), \dots, A_6=circ(a_{16},a_{17}, a_{18}),  \\
A_7=revcirc(a_{19},a_{20}, a_{21}), \dots  A_{12}=revcirc(a_{34},a_{35},a_{36})
\end{aligned}
\end{equation}

 \begin{equation}\label{three}
\begin{aligned}
A_1=circ(a_1,a_2,a_3),\dots, A_{3}=circ(a_{7},a_{8}, a_{9}),\\
A_4=revcirc(a_{10},a_{11}, a_{12}), \dots, A_{6}=revcirc(a_{16},a_{17}, a_{18}),\\
A_7=circ(a_{19},a_{20}, a_{21}),\dots, A_{9}=circ(a_{25},a_{26}, a_{27}),\\
A_{10}=revcirc(a_{28},a_{29}, a_{30}), \dots, A_{12}=revcirc(a_{34},a_{35}, a_{36}).
\end{aligned}
\end{equation}

 \begin{equation}\label{four}
\begin{aligned}
A_1=revcirc(a_1,a_2,a_3), A_{2}=revcirc(a_4,a_5,a_6),\\
 A_{3}=circ(a_{7},a_{8}, a_{9}), A_4=circ(a_{10},a_{11}, a_{12}),\\
 A_{5}=revcirc(a_{13},a_{14}, a_{15}), A_{6}=revcirc(a_{16},a_{17}, a_{18}),\\
A_7=circ(a_{19},a_{20}, a_{21}), A_{8}=circ(a_{22},a_{23}, a_{24}), \\
A_{9}=revcirc(a_{25},a_{26}, a_{27}), A_{10}=revcirc(a_{28},a_{29}, a_{30}),\\
 A_{11}=circ(a_{31},a_{32}, a_{33}), A_{12}=circ(a_{34},a_{35}, a_{36}).\\
\end{aligned}
\end{equation}

 \begin{equation}\label{six}
\begin{aligned}
A_1=revcirc(a_1,a_2,a_3), A_{2}=circ(a_4,a_5,a_6),\\
 A_{3}=revcirc(a_{7},a_{8}, a_{9}), A_4=circ(a_{10},a_{11}, a_{12}),\\
 A_{5}=revcirc(a_{13},a_{14}, a_{15}), A_{6}=circ(a_{16},a_{17}, a_{18}),\\
A_7=revcirc(a_{19},a_{20}, a_{21}), A_{8}=circ(a_{22},a_{23}, a_{24}), \\
A_{9}=revcirc(a_{25},a_{26}, a_{27}), A_{10}=circ(a_{28},a_{29}, a_{30}),\\
 A_{11}=revcirc(a_{31},a_{32}, a_{33}), A_{12}=circ(a_{34},a_{35}, a_{36}).\\
\end{aligned}
\end{equation}

 \begin{equation}\label{seven}
\begin{aligned}
A_1=circ(a_1,a_2,a_3), A_{2}=revcirc(a_4,a_5,a_6),\\
 A_{3}=circ(a_{7},a_{8}, a_{9}), A_4=revcirc(a_{10},a_{11}, a_{12}),\\
 A_{5}=circ(a_{13},a_{14}, a_{15}), A_{6}=revcirc(a_{16},a_{17}, a_{18}),\\
A_7=circ(a_{19},a_{20}, a_{21}), A_{8}=revcirc(a_{22},a_{23}, a_{24}), \\
A_{9}=circ(a_{25},a_{26}, a_{27}), A_{10}=revcirc(a_{28},a_{29}, a_{30}),\\
 A_{11}=circ(a_{31},a_{32}, a_{33}), A_{12}=revcirc(a_{34},a_{35}, a_{36}).\\
\end{aligned}
\end{equation}

 \begin{equation}\label{eight}
\begin{aligned}
A_1=circ(a_1,a_2,a_3), A_{2}=circ(a_4,a_5,a_6),\\
 A_{3}=revcirc(a_{7},a_{8}, a_{9}), A_4=revcirc(a_{10},a_{11}, a_{12}),\\
 A_{5}=circ(a_{13},a_{14}, a_{15}), A_{6}=circ(a_{16},a_{17}, a_{18}),\\
A_7=revcirc(a_{19},a_{20}, a_{21}), A_{8}=revcirc(a_{22},a_{23}, a_{24}),\\
A_{9}=circ(a_{25},a_{26}, a_{27}), A_{10}=circ(a_{28},a_{29}, a_{30}),\\
 A_{11}=revcirc(a_{31},a_{32}, a_{33}), A_{12}=revcirc(a_{34},a_{35}, a_{36}).\\
\end{aligned}
\end{equation}

We only tried limited cases of these twelve matrices since there are a lot of choices for this. We remark that, one can also consider some other combinations of these twelve $A_i$ matrices.

\subsection{The Dihedral Group $D_{12}$}
Here, we consider the dihedral group of order 12 $$D_{12}=\langle a,b \ | \ b^6=a^2=1, ab=b^{-1}a \rangle$$ and 2 different ordering of the group elements. \\

\textbf{Case 1:}\\
 Let $v_1=\sum_{i=0}^5 \sum_{j=0}^1 \alpha_{a^ib^j}a^ib^j \in M_3(\mathbb{F}_2)D_{12},$ and consider the
ordering of the elements of the group as
 $$1, b^{5}, b^{4}, \dots, b^{1}, a, ab, \dots, ab^{6}, $$  then\\
\begin{equation}
\tau_3(v_1)=\begin{pmatrix}
A&B\\
B^T&A^T
\end{pmatrix},
\end{equation}
where $A=CIRC(A_1,A_2,A_3,\dots,A_6), B=CIRC(A_{7},A_{11},A_{12},\dots,A_{12})$. We define the following generator matrices;

$$ \mathcal{G}_1^{1}=[I_{36} \ | \ \tau_3(v_1)], $$ where $A_i$'s come from Equation \ref{one}.

$$ \mathcal{G}_1^{2}=[I_{36} \ | \ \tau_3(v_1)], $$ where $A_i$'s come from Equation \ref{two}.

\textbf{Case 2:}\\
Let $v_2=\sum_{i=0}^5 \sum_{j=0}^1 \alpha_{a^ib^j}a^ib^j \in M_3(\mathbb{F}_2)D_{12}$  and consider the
ordering of the elements of the group as
 $$1, b, b^{2}, \dots, b^{5}, a, ab, \dots, ab^{5}, $$ then\\
\begin{equation}
\sigma_3(v_2)=\begin{pmatrix}
A&B\\
B&A
\end{pmatrix},
\end{equation}
where $ A=CIRC(A_{1}, A_{2}, \dots, A_{6} ),$ $ B=REVCIRC(A_{7}, A_{11}, \dots, A_{12} ) $.  We define the following generator matrices;\\

$$ \mathcal{G}_2^{1}=[I_{36} \ | \ \tau_3(v_2)], $$ where $A_i$'s come from Equation \ref{three}.

$$ \mathcal{G}_2^{2}=[I_{36} \ | \ \tau_3(v_2)], $$  where $A_i$'s come from Equation \ref{four}.

$$ \mathcal{G}_2^{3}=[I_{36} \ | \ \tau_3(v_2)], $$ where $A_i$'s come from Equation \ref{five}.

$$ \mathcal{G}_2^{4}=[I_{36} \ | \ \tau_3(v_2)], $$ where $A_i$'s come from Equation \ref{six}.

$$ \mathcal{G}_2^{5}=[I_{36} \ | \ \tau_3(v_2)], $$ where $A_i$'s come from Equation \ref{seven}.

$$ \mathcal{G}_2^{6}=[I_{36} \ | \ \tau_3(v_2)], $$ where $A_i$'s come from Equation \ref{eight}.

\subsection{The Cyclic Group $C_{12}$}
Let $C_{12}$ be the  cyclic group of order 12 and $v_3 \in M_3(\mathbb{F}_2)C_{12}$. We consider the following two matrix representations; \\

\textbf{Case 1:}
\begin{equation}
\sigma_3(v_3)=\begin{pmatrix}
A&B\\
B'&A
\end{pmatrix},
\end{equation}
where $A=CIRC(A_1,A_2,A_3,\dots,A_6), B=CIRC(A_{7},A_{11},A_{12},\dots,A_{12})$ and $B'=CIRC(A_{12},A_{7},A_{8},\dots,A_{11})$.
 We define the following generator matrices;

$$ \mathcal{G}_3^{1}=[I_{36} \ | \ \tau_3(v_3)], $$  where $A_i$'s come from Equation \ref{nine}.

$$ \mathcal{G}_3^{2}=[I_{36} \ | \ \tau_3(v_3)], $$  where $A_i$'s come from Equation \ref{two}.

$$ \mathcal{G}_3^{3}=[I_{36} \ | \ \tau_3(v_3)], $$ where $A_i$'s come from Equation \ref{five}.

$$ \mathcal{G}_3^{4}=[I_{36} \ | \ \tau_3(v_3)],$$ where $A_i$'s come from Equation \ref{one}.

\textbf{Case 2:}
\begin{equation}
\sigma_3(v_4)=\begin{pmatrix}
A&B&C&D\\
D'&A&B&C\\
C'&D'&A&B\\
B'&C'&D'&A
\end{pmatrix},
\end{equation}
where $A=CIRC(A_1,A_2,A_3), B=CIRC(A_{4},A_{5},A_{6}),$ $B'=CIRC(A_{6},A_{4},A_{5})$,
$C=CIRC(A_{7}, A_{8}, A_{9}), C'=CIRC(A_{9}, A_{7}, A_{8}),$ $D=CIRC(A_{10},A_{11},A_{12})$ and
$D=CIRC(A_{12},A_{10},A_{11}).$  We define the following generator matrices;

$$ \mathcal{G}_4^{1}=[I_{36} \ | \ \tau_3(v_4)], $$  where $A_i$'s come from Equation \ref{three}.

 $$ \mathcal{G}_4^{2}=[I_{36} \ | \ \tau_3(v_4)], $$  where $A_i$'s come from Equation \ref{two}.

 $$ \mathcal{G}_4^{3}=[I_{36} \ | \ \tau_3(v_4)], $$  where $A_i$'s come from Equation \ref{nine}.

\subsection{The Group $C_{6}\times C_{2}$}

Let $C_{6}\times C_{2}$ be the cross product of cyclic groups of order 6 and 2. Then for  $v_5 \in M_3(\mathbb{F}_2)(C_{6}\times C_{2})_{12},$ we have\\
\begin{equation}
\sigma_3(v_5)=\begin{pmatrix}
A&B\\
B&A
\end{pmatrix},
\end{equation}
where $A=CIRC(A_1,A_2,A_3,\dots,A_6), B=CIRC(A_{7},A_{11},A_{12},\dots,A_{12})$.  We now define the following generator matrices;

 $$ \mathcal{G}_5^{1}=[I_{36} \ | \ \tau_3(v_5)], $$  where $A_i$'s come from Equation \ref{one}.

 $$ \mathcal{G}_5^{2}=[I_{36} \ | \ \tau_3(v_5)], $$  where $A_i$'s come from Equation \ref{three}.

 $$ \mathcal{G}_5^{3}=[I_{36} \ | \ \tau_3(v_5)], $$ where $A_i$'s come from Equation \ref{five}.

\subsection{The Group $C_{3}\times C_{4}$}

Let $C_{3}\times C_{4}$ be the cross product of cyclic groups of order 3 and 4. Then for  $v_6 \in M_3(\mathbb{F}_2)(C_{3}\times C_{4})_{12},$ we have\\
\begin{equation}
\sigma_3(v_6)=\begin{pmatrix}
A&B&C&D\\
D&A&B&C\\
C&D&A&B\\
B&C&D&A
\end{pmatrix},
\end{equation}
where $A=CIRC(A_1,A_2, A_3), B=CIRC(A_{4},A_{5}, A_6),  C=CIRC(A_{7},A_{8}, A_9), D=CIRC(A_{10},A_{11},A_{12})$. We now define the following generator matrices;

$$ \mathcal{G}_6^{1}=[I_{36} \ | \ \tau_3(v_6)], $$ where $A_i$'s come from Equation \ref{seven}.

 $$ \mathcal{G}_6^{2}=[I_{36} \ | \ \tau_3(v_6)], $$ where $A_i$'s come from Equation \ref{four}.

\subsection{The Alternating Group $A_{4}$}

Let $A_{4}$ be the alternating group of order 12 and $v_7 \in M_3(\mathbb{F}_2)(A_{4}),$ then we have the following generator matrix from \cite{Dougherty6}\\

\begin{equation}
\sigma_3(v_7)=\left(\begin{smallmatrix}
A_{1}& A_{2}& A_{3}& A_{4}& A_{5}& A_{6}& A_{7}& A_{8}& A_{9}& A_{10}& A_{11}& A_{12}\\
A_{3}& A_{1}& A_{2}& A_{12}& A_{10}& A_{11}& A_{6}& A_{4}& A_{5}& A_{9}& A_{7}& A_{8}\\
A_{2}& A_{3}& A_{1}& A_{8}& A_{9}& A_{7}& A_{11}& A_{12}& A_{10}& A_{5}& A_{6}& A_{4}\\
A_{4}& A_{5}& A_{6}& A_{1}& A_{2}& A_{3}& A_{10}& A_{11}& A_{12}& A_{7}& A_{8}& A_{9}\\
A_{12}& A_{10}& A_{11}& A_{3}& A_{1}& A_{2}& A_{9}& A_{7}& A_{8}& A_{6}& A_{4}& A_{5}\\
A_{8}& A_{9}& A_{7}& A_{2}& A_{3}& A_{1}& A_{5}& A_{6}& A_{4}& A_{11}& A_{12}& A_{10}\\
A_{7}& A_{8}& A_{9}& A_{10}& A_{11}& A_{12}& A_{1}& A_{2}& A_{3}& A_{4}& A_{5}& A_{6}\\
A_{6}& A_{4}& A_{5}& A_{9}& A_{7}& A_{8}& A_{3}& A_{1}& A_{2}& A_{12}& A_{10}& A_{11}\\
A_{11}& A_{12}& A_{10}& A_{5}& A_{6}& A_{4}& A_{2}& A_{3}& A_{1}& A_{8}& A_{9}& A_{7}\\
A_{10}& A_{11}& A_{12}& A_{7}& A_{8}& A_{9}& A_{4}& A_{5}& A_{6}& A_{1}& A_{2}& A_{3}\\
A_{9}& A_{7}& A_{6}& A_{6}& A_{4}& A_{5}& A_{12}& A_{10}& A_{11}& A_{3}& A_{1}& A_{2}\\
A_{5}& A_{6}& A_{4}& A_{10}& A_{12}& A_{10}& A_{8}& A_{9}& A_{7}& A_{2}& A_{3}& A_{1}\\
\end{smallmatrix}\right),
\end{equation}
We now define the following generator matrices;

$$ \mathcal{G}_7^{1}=[I_{36} \ | \ \tau_3(v_7)], $$ where $A_i$'s come from Equation \ref{three}.

$$ \mathcal{G}_7^{2}=[I_{36} \ | \ \tau_3(v_7)],  $$ where $A_i$'s come from Equation \ref{eight}.

$$ \mathcal{G}_7^{3}=[I_{36} \ | \ \tau_3(v_7)],  $$ where $A_i$'s come from Equation \ref{seven}.

 $$ \mathcal{G}_7^{4}=[I_{36} \ | \ \tau_3(v_7)], $$ where $A_i$'s come from Equation \ref{four}.

 $$ \mathcal{G}_7^{5}=[I_{36} \ | \ \tau_3(v_7)],  $$ where $A_i$'s come from Equation \ref{five}.

\subsection{Dicyclic Group $Dic_{12}$}

We consider the dicyclic group of order 12
$$Dic_{12}= \langle x,y | x^6=1, ~~y^2=x^3, x^y=x^{-1}\rangle.$$
Let $v_8 \in M_3(\mathbb{F}_2)(Dic_{12}),$ we have the following generator matrix from \cite{Dougherty7}\\

\begin{equation}
\sigma_3(v_8)=\begin{pmatrix}
A&B\\
C&A
\end{pmatrix},
\end{equation}
where $ A=CIRC(A_{1}, A_{2}, \dots, A_{6} ),$ $ B=REVCIRC(A_{7}, A_{11}, \dots, A_{12} ),$ and $C=REVCIRC(A_{10}, A_{11}, A_{12},A_{7}, A_{8}, A_{9} ).$  We define the following generator matrices;\\

$$ \mathcal{G}_8^{1}=[I_{36} \ | \ \tau_3(v_8)], $$ where $A_i$'s come from Equation \ref{two}.

$$ \mathcal{G}_8^{2}=[I_{36} \ | \ \tau_3(v_8)], $$  where $A_i$'s come from Equation \ref{five}.

$$ \mathcal{G}_8^{3}=[I_{36} \ | \ \tau_3(v_8)], $$ where $A_i$'s come from Equation \ref{eight}.

\section{Virus Optimization Algorithm for Self-Dual Codes}
Metaheuristic algorithms have been very useful and powerful tool to simulate real-world problems that were previously difficult or impossible to solve \cite{Ustun}.  For algebraic coding theory,  optimization algorithms were used  only for  the minimum distance problem, for example \cite{Shenoy,Bland1, Bland2, Bouzkraoui}.
\begin{figure*}[h!]
\centering
\includegraphics[width=110mm]{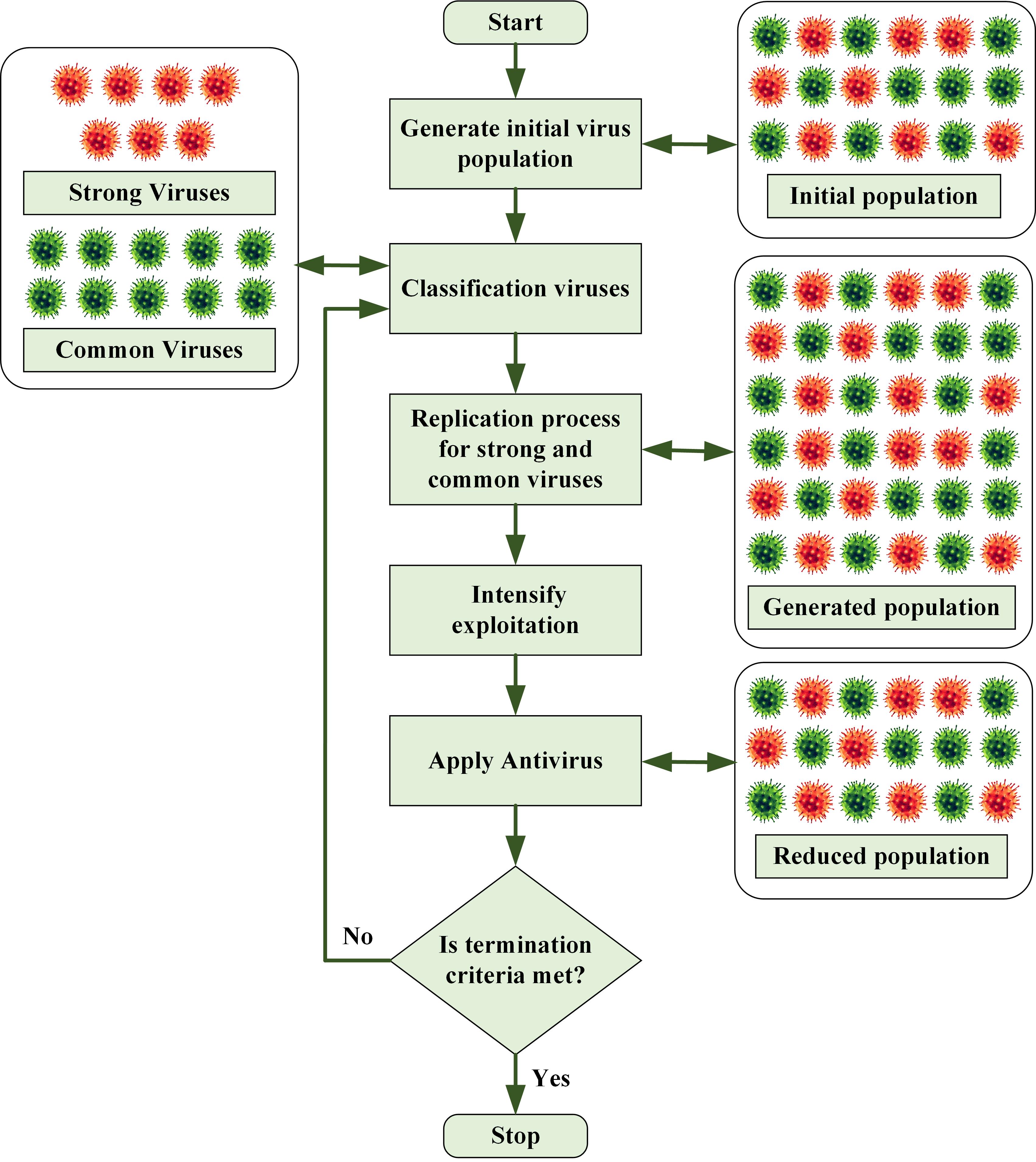}
\caption{Flowchart of VOA  \label{VOA}}
\end{figure*}

 But recently in \cite{Korban1}, one of the oldest and well-known optimization algorithm, genetic algorithm, was applied for the first time to the problem of finding binary self-dual codes.
 VOA which is one of nature-inspired optimization approaches using iteratively population-based, mimics the behaviour of the viruses assaulting a living cells. In the VOA, the count of the viruses in the population increase at every replication process and a process called as antivirus  is applied to population by immune system to prevent the population of virus uncontrollably growth. In the algorithm, there are two types of viruses called as strong and common viruses respectively for establishing the balance between its exploration and exploitation abilities. The flowchart in Figure \ref{VOA} gives the steps of the VOA. The algorithm includes three major steps that are named as initialization, replication and updating/maintenance. At initial process, the first population of virus is generated by using the VOA's parameters specified by user.  The values of the objective function computed for each of the viruses are ordered and strong and common viruses in the population are chosen according to these values. In the replication process, new viruses are generated by taking account of the strong and common viruses. The number of viruses in population is controlled by using antivirus mechanism provided by Maintenance/updating. In the VOA, unless the performance of the population enhances, exploitation is intensified and antivirus is applied to population. If termination criteria has not been attained, the counter of replication is increased by one and then other replication processes are continued; otherwise the algorithm stops.

\begin{figure*}[h!]
\centering
\includegraphics[width=130mm]{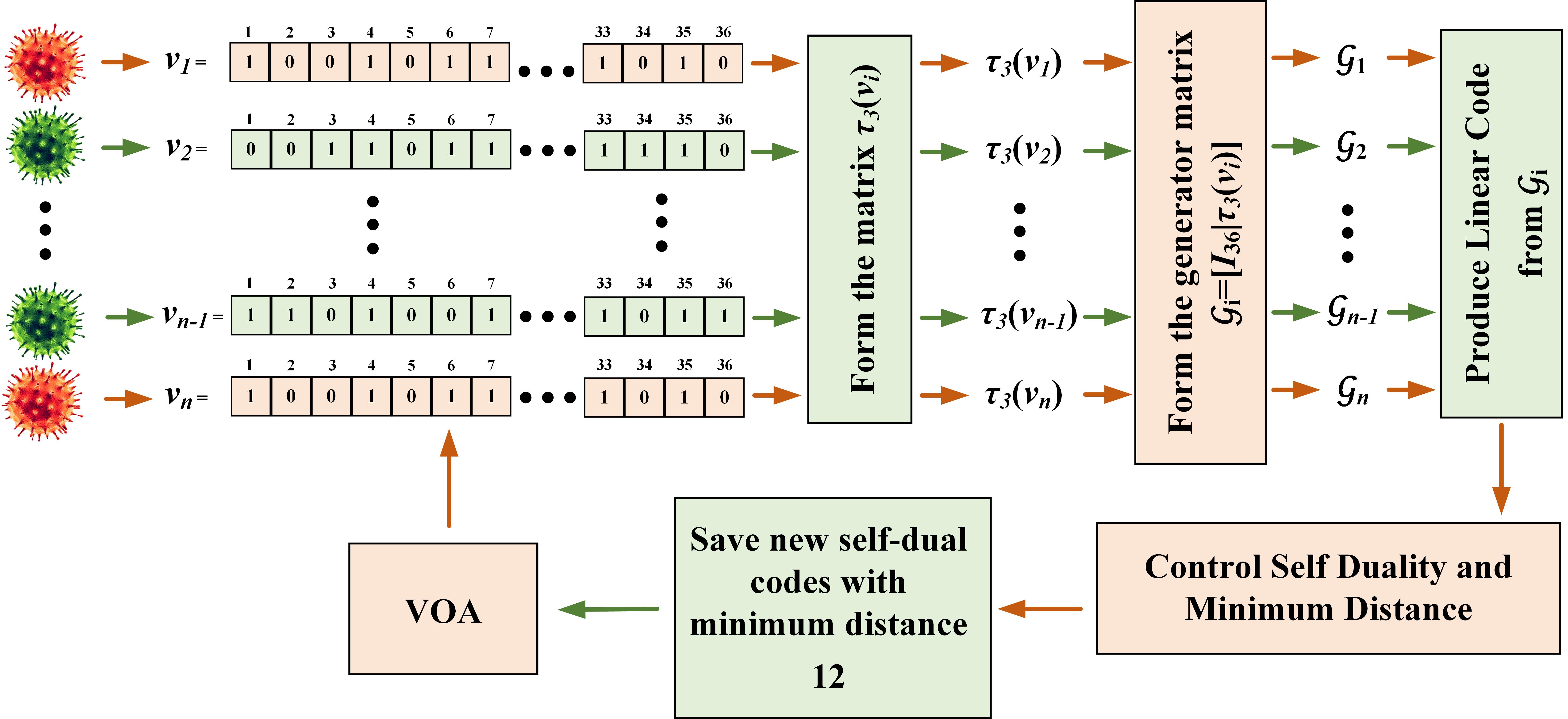}
\caption{Flowchart of Code Generation Process by VOA \label{VOA2}}
\end{figure*}

An application of VOA to the problem of finding binary self-dual codes is given in Figure \ref{VOA2}. For the considered problem, each of the members in the population are represented by a vector with length 36 and these vectors are elements of $\mathbb{F}_2^{36}$. These vectors form the first row of the matrices  $\tau_k(v)$. Then, the  linear codes are generated from the generator matrices of the form $[I_{36} \ | \ \tau_3(v_i)]$.
If a generated code is self-dual and has a minimum distance 12 then it is saved. The VOA generates new vectors by applying the process of replication to the old strong and common viruses, until the algorithm's termination criteria value.

\begin{figure*}[h!]
\centering
\includegraphics[width=100mm]{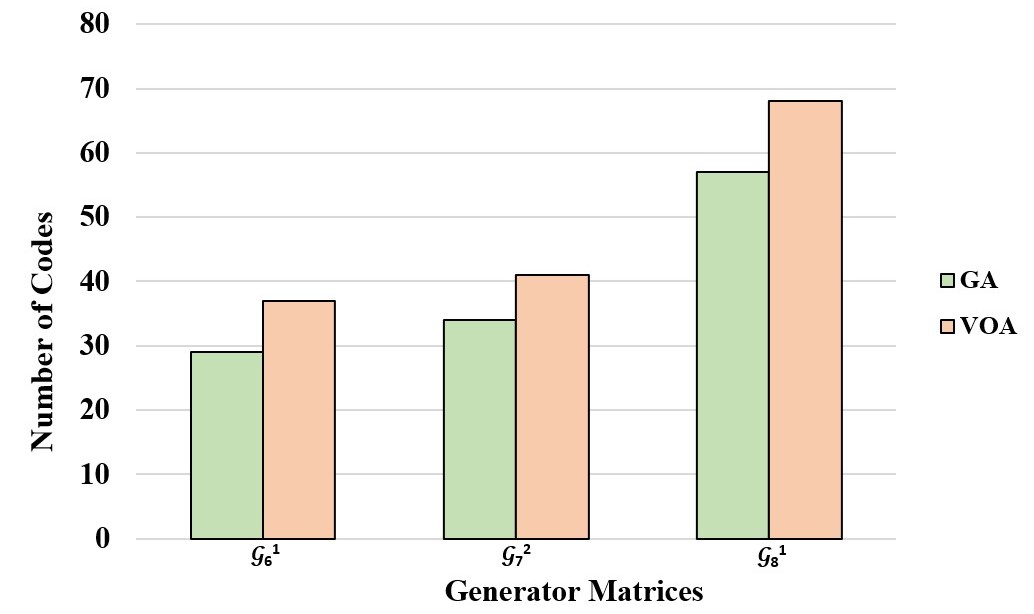}
\caption{Comparasion of VOA and GA \label{Comp2}}
\end{figure*}

It was shown in \cite{Liang} that the VOA outperforms the GA in continues problems. Therefore it is natural to compare these two algorithms for the considered problem by performing a search for extremal binary self-dual codes that have generator matrices of the  $[I_{36} \ | \ \tau_3(v_i)]$ form.  The comparison of  GA and VOA in terms of number of codes found is given in the Figure \ref{Comp2}. Calculations for the comparison of GA and VOA were done only for three generator matrices; $\mathcal{G}_6^{1}, \mathcal{G}_7^{2}$ and $\mathcal{G}_8^{1}$. We used the same software - Magma \cite{MAGMA}, for the two approaches and  algorithms were run on a workstation with Intel Xeon 4.0 GHz processor and 64 GByte RAM. The size of the population and number of iteration  for both GA and VOA are 500 and 100, respectively.

\section{Computational Results}

In this section, we employ the generator matrices defined in Section~3 and search for binary $[72,36,12]$ self-dual codes.

The possible weight enumerators for a Type~I $[72,36,12]$ codes are as follows (\cite{Dougherty4}):
$$W_{72,1}=1+2\beta y^{12}+(8640-64\gamma)y^{14}+(124281-24\beta+384\gamma)y^{16}+\dots$$
$$W_{72,2}=1+2 \beta y^{12}+(7616-64 \gamma)y^{14}+(134521-24 \beta+384 \gamma)y^{16}+\dots$$
where $\beta$ and $\gamma$ are parameters.
The possible weight enumerators for Type~II $[72,36,12]$ codes are (\cite{Dougherty4}):
$$1+(4398+\alpha)y^{12}+(197073-12\alpha)y^{16}+(18396972+66\alpha)y^{20}+\dots $$
where $\alpha$ is a parameter. For an up-to-date list of all known Type~I and Type~II binary self-dual codes with parameters $[72,36,12]$ please see \cite{selfdual72}.

We only list codes with parameters in their weight enumerators that were not known in the literature before. All the upcoming computational results were obtained by performing searches in the software package MAGMA (\cite{MAGMA}).

\begin{table}[h!]\label{Type1}
\caption{New Binary Type I  Codes of Length 72}
\resizebox{0.65\textwidth}{!}{\begin{minipage}{\textwidth}
\centering
\begin{tabular}{cccccccc}
\hline
& Generator Matrix      & Type       & $r_A$                                   & $r_B$                                   & $\gamma$ & $\beta$ & $|Aut(C_i)|$ \\ \hline
$C_{1}$  &$\mathcal{G}_1^{1}$& $W_{72,1}$ & $(0, 1, 0, 1, 1, 1, 1, 1, 0, 1, 1, 1, 0, 1, 0, 0, 0, 0)$ & $(  1, 0, 1, 1, 1, 1, 0, 1, 0, 0, 1, 0, 1, 1, 1, 0, 1, 0)$ & $0$      & $129$   & $72$         \\ \hline
$C_{2}$ &$\mathcal{G}_1^{2}$& $W_{72,1}$ & $(1, 1, 0, 1, 0, 0, 0, 0, 0, 1, 0, 0, 1, 1, 0, 1, 0, 1)$ & $( 1, 0, 0, 0, 0, 1, 0, 1, 0, 1, 0, 1, 1, 1, 0, 0, 1, 1)$ & $36$      & $483$   & $72$         \\ \hline
$C_{3}$ &$\mathcal{G}_2^{1}$& $W_{72,1}$ & $( 1, 0, 0, 1, 0, 0, 1, 1, 1, 1, 0, 1, 1, 1, 0, 0, 0, 0)$& $( 0, 0, 0, 1, 0, 0, 1, 1, 0, 0, 0, 0, 0, 0, 0, 1, 0, 0)$ & $0$      & $236$   & $12$      \\ \hline
$C_{4}$  &$\mathcal{G}_2^{2}$& $W_{72,1}$ & $(0, 0, 1, 0, 0, 1, 1, 1, 1, 1, 0, 1, 0, 1, 0, 0, 0, 1)$ & $( 1, 0, 1, 0, 1, 0, 1, 0, 0, 1, 1, 0, 0, 0, 1, 0, 0, 1)$ & $6$      & $341$   & $12$      \\ \hline
$C_{5}$ &$\mathcal{G}_2^{1}$& $W_{72,1}$ & $(0, 0, 1, 0, 0, 1, 1, 0, 1, 1, 0, 0, 1, 0, 0, 0, 1, 1)$ & $( 0, 1, 1, 1, 0, 0, 1, 1, 1, 0, 0, 1, 0, 0, 1, 1, 0, 0)$ & $6$      & $353$   & $12$      \\ \hline
$C_{6}$  &$\mathcal{G}_2^{1}$& $W_{72,1}$ & $(0, 1, 1, 1, 0, 0, 0, 1, 1, 0, 1, 1, 0, 1, 1, 1, 1, 0)$ & $( 0, 0, 1, 0, 0, 1, 0, 1, 1, 0, 1, 0, 0, 1, 0, 0, 0, 0)$ & $12$     & $269$   & $12$       \\ \hline
$C_{7}$ &$\mathcal{G}_2^{1}$& $W_{72,1}$ & $(1, 0, 0, 0, 1, 1, 1, 0, 0, 1, 0, 0, 1, 0, 0, 0, 1, 0)$ & $(0, 1, 1, 0, 0, 1, 0, 0, 1, 0, 0, 0, 0, 0, 1, 0, 0, 1)$  & $12$     & $285$   & $12$       \\ \hline
$C_{8}$ &$\mathcal{G}_2^{1}$& $W_{72,1}$ & $(1, 0, 1, 0, 1, 1, 0, 0, 1, 1, 0, 0, 1, 0, 0, 1, 0, 1)$ & $( 1, 1, 1, 1, 0, 1, 1, 1, 0, 1, 0, 1, 0, 1, 0, 1, 0, 1)$ & $12$     & $314$   & $12$       \\ \hline
$C_{9}$  &$\mathcal{G}_2^{1}$& $W_{72,1}$ & $(0, 1, 0, 0, 0, 1, 0, 0, 1, 1, 0, 0, 0, 1, 1, 1, 0, 0)$ & $(0, 1, 1, 1, 0, 0, 1, 1, 1, 1, 0, 1, 0, 1, 0, 1, 0, 0)$  & $18$     & $279$   & $12$       \\ \hline
$C_{10}$ &$\mathcal{G}_2^{1}$& $W_{72,1}$ & $(1, 1, 1, 1, 1, 0, 1, 1, 0, 0, 0, 0, 1, 0, 0, 0, 1, 0)$ & $( 0, 1, 1, 1, 1, 0, 1, 1, 1, 1, 1, 0, 1, 1, 1, 0, 0, 0)$ & $36$     & $434$   & $12$       \\ \hline
$C_{11}$  &$\mathcal{G}_2^{1}$& $W_{72,1}$ & $(0, 1, 0, 0, 0, 0, 0, 1, 1, 0, 0, 0, 1, 1, 1, 0, 1, 1)$ & $(0, 1, 0, 0, 0, 0, 1, 0, 1, 1, 0, 1, 1, 1, 1, 0, 0, 1)$  & $36$     & $458$   & $12$         \\ \hline
$C_{12}$  &$\mathcal{G}_2^{1}$& $W_{72,1}$ & $(0, 1, 1, 1, 0, 1, 1, 1, 1, 1, 0, 0, 1, 0, 0, 0, 0, 0)$ & $(1, 1, 1, 0, 0, 0, 1, 1, 0, 0, 0, 0, 1, 0, 0, 1, 1, 0)$  & $36$     & $470$   & $12$         \\ \hline
$C_{13}$ &$\mathcal{G}_2^{1}$& $W_{72,1}$ & $(0, 0, 0, 0, 1, 0, 1, 1, 1, 0, 0, 0, 0, 1, 0, 0, 1, 0)$ & $(1, 0, 1, 1, 1, 1, 1, 0, 1, 0, 0, 0, 1, 0, 1, 1, 1, 0)$  & $30$     & $495$   & $12$         \\ \hline
$C_{14}$ &$\mathcal{G}_2^{2}$& $W_{72,1}$ & $(0, 0, 0, 1, 0, 0, 1, 0, 1, 1, 0, 1, 1, 0, 1, 0, 0, 0)$ & $(0, 0, 1, 1, 0, 1, 0, 0, 1, 1, 1, 1, 0, 1, 1, 0, 0, 1)$  & $6$      & $281$   & $12$         \\ \hline
$C_{15}$ &$\mathcal{G}_2^{2}$& $W_{72,1}$ & $(0, 1, 0, 0, 1, 0, 0, 0, 1, 1, 0, 0, 1, 1, 1, 1, 1, 0)$ & $(1, 0, 0, 0, 1, 0, 1, 1, 0, 0, 1, 0, 0, 0, 1, 1, 1, 0)$  & $6$      & $252$   & $12$         \\ \hline
$C_{16}$ &$\mathcal{G}_2^{2}$& $W_{72,1}$ & $(1, 0, 1, 0, 0, 1, 1, 1, 0, 0, 0, 0, 0, 1, 1, 1, 0, 1)$ & $(0, 0, 0, 0, 1, 0, 0, 0, 1, 1, 1, 0, 1, 0, 0, 1, 1, 1)$  & $12$     & $357$   & $12$         \\ \hline
$C_{17}$ &$\mathcal{G}_2^{2}$& $W_{72,1}$ & $(0, 1, 1, 0, 0, 0, 1, 1, 0, 0, 0, 1, 1, 1, 1, 0, 0, 1)$ & $(0, 1, 1, 1, 1, 0, 1, 1, 1, 1, 0, 1, 1, 1, 1, 0, 0, 0)$  & $36$     & $506$   & $12$   \\ \hline
$C_{18}$ &$\mathcal{G}_2^{4}$& $W_{72,1}$ & $(0, 0, 1, 1, 0, 0, 1, 1, 0, 1, 1, 0, 0, 0, 0, 0, 0, 1)$ & $(0, 1, 0, 1, 1, 0, 1, 0, 1, 0, 1, 1, 0, 1, 0, 1, 1, 0 )$ & $18$     & $324$   & $26$   \\ \hline
$C_{19}$ &$\mathcal{G}_2^{6}$& $W_{72,1}$ & $(0, 0, 1, 0, 1, 0, 1, 0, 0, 0, 0, 1, 1, 0, 1, 0, 0, 1)$ & $(0, 1, 0, 0, 1, 0, 1, 1, 0, 1, 0, 0, 0, 1, 0, 0, 0, 0)$  & $0$      & $359$   & $12$  \\ \hline
$C_{20}$ &$\mathcal{G}_2^{6}$& $W_{72,1}$ & $(1, 0, 0, 0, 0, 1, 0, 1, 1, 1, 0, 0, 1, 0, 0, 1, 1, 0)$ & $(1, 1, 0, 0, 1, 1, 1, 0, 1, 1, 1, 1, 1, 1, 0, 0, 1, 1)$  & $6$      & $276$   & $12$  \\ \hline
$C_{21}$ &$\mathcal{G}_3^{1}$ & $W_{72,1}$& $(1, 1, 1, 0, 1, 0, 0, 1, 0, 0, 1, 1, 0, 0, 1, 0, 0, 1)$ & $(1, 0, 0, 1, 0, 1, 1, 0, 0, 1, 0, 0, 0, 1, 1, 0, 1, 0)$  & $36$     & $504$   & $72$  \\ \hline
$C_{22}$ &$\mathcal{G}_3^{1}$& $W_{72,1}$ & $(1, 0, 1, 1, 0, 1, 1, 0, 0, 1, 1, 0, 0, 0, 0, 1, 0, 1)$ & $(0, 0, 0, 1, 1, 1, 1, 0, 0, 0, 1, 1, 0, 1, 0, 1, 0, 0)$  & $36$     & $552$   & $72$  \\ \hline
$C_{23}$ &$\mathcal{G}_3^{2}$& $W_{72,1}$ & $(0, 1, 0, 1, 0, 1, 1, 0, 0, 1, 0, 1, 0, 0, 1, 1, 0, 1)$ & $(1, 0, 1, 0, 1, 0, 1, 0, 1, 1, 0, 0, 0, 0, 0, 1, 1, 0)$  & $0$      & $426$   & $72$  \\ \hline
$C_{24}$ &$\mathcal{G}_3^{2}$& $W_{72,1}$ & $(0, 0, 1, 1, 1, 1, 0, 1, 1, 1, 1, 1, 1, 1, 1, 0, 0, 0)$ & $(0, 0, 0, 0, 1, 1, 1, 1, 0, 0, 0, 1, 1, 1, 0, 1, 0, 1)$  & $36$     & $456$   & $72$   \\ \hline
$C_{25}$ &$\mathcal{G}_3^{2}$& $W_{72,1}$ & $(1, 1, 1, 0, 0, 0, 1, 0, 1, 0, 0, 0, 1, 0, 0, 1, 1, 1)$ & $(1, 1, 0, 1, 0, 0, 1, 1, 0, 1, 0, 1, 1, 1, 0, 1, 1, 1)$  & $36$     & $633$   & $72$   \\ \hline
$C_{26}$ &$\mathcal{G}_5^{1}$& $W_{72,1}$ & $(1, 0, 0, 1, 0, 1, 1, 1, 1, 1, 0, 1, 1, 0, 0, 1, 1, 0)$ & $(1, 1, 0, 1, 1, 0, 0, 1, 0, 0, 0, 0, 1, 1, 1, 1, 0, 1)$  & $0$      & $69$    & $136$   \\ \hline
$C_{27}$ &$\mathcal{G}_5^{2}$& $W_{72,1}$ & $(0, 0, 1, 1, 0, 0, 1, 1, 0, 0, 1, 0, 0, 0, 1, 1, 0, 1)$ & $(1, 0, 1, 0, 0, 0, 1, 0, 1, 1, 1, 0, 1, 1, 0, 0, 0, 1)$  & $12$     & $321$   & $12$   \\ \hline
$C_{28}$ &$\mathcal{G}_5^{1}$& $W_{72,1}$ & $(1, 1, 0, 1, 1, 1, 1, 1, 1, 0, 0, 0, 0, 0, 1, 0, 0, 1)$ & $(1, 0, 1, 1, 0, 1, 0, 1, 1, 1, 1, 1, 0, 0, 0, 0, 1, 1)$  & $0$      & $327$   & $72$   \\ \hline
$C_{29}$ &$\mathcal{G}_5^{3}$& $W_{72,1}$ & $(1, 1, 1, 1, 0, 0, 1, 0, 1, 0, 0, 1, 0, 1, 0, 0, 1, 1)$ & $(1, 0, 1, 0, 0, 0, 1, 1, 1, 1, 0, 1, 1, 1, 0, 1, 1, 0)$  & $0$      & $75$    & $36$    \\ \hline
$C_{30}$ &$\mathcal{G}_8^{1}$& $W_{72,1}$ & $(0, 0, 1, 1, 0, 1, 0, 1, 0, 1, 1, 0, 1, 0, 1, 1, 0, 1)$ & $(1, 1, 1, 0, 0, 1, 1, 1, 1, 1, 0, 1, 0, 0, 0, 1, 0, 1)$  & $36$      & $468$    & $72$    \\ \hline
\end{tabular}
\end{minipage}}
\end{table}

\begin{table}[h!]\label{Type1ABCD}
\caption{New Binary Type I  Codes of Length 72}
\resizebox{0.60\textwidth}{!}{\begin{minipage}{\textwidth}
\centering
\begin{tabular}{cccccccccc}
\hline
& Generator Matrix       & Type       & $r_A$       & $r_B$ & $r_C$       & $r_D$                                   & $\gamma$ & $\beta$ & $|Aut(C_i)|$ \\ \hline
$C_{31}$ &$\mathcal{G}_4^{1}$ & $W_{72,1}$ & $(1, 0, 1, 1, 0, 1, 1, 0, 1)$ & $(0, 1, 0, 1, 0, 0, 1, 0, 0)$ & $(1, 1, 0, 1, 1, 0, 1, 0, 1)$ & $(0, 0, 1, 1, 1, 1, 1,1, 0)$ & $0$   & $99$   &$36$  \\ \hline
$C_{32}$  &$\mathcal{G}_6^{1}$& $W_{72,1}$ & $(1, 1, 0, 1, 1, 1, 1, 1, 1)$ & $(0, 1, 0, 1, 1, 0, 1, 0, 1)$ & $(0, 0, 0, 0, 0, 0, 1, 1, 0)$ & $(1, 1, 0, 1, 1, 1, 0, 0, 1)$& $12$  & $267$  &$24$   \\ \hline
$C_{33}$ &$\mathcal{G}_6^{2}$& $W_{72,1}$ & $(1, 1, 1, 1, 1, 1, 1, 0, 0)$ & $( 0, 0, 0, 1, 0, 0, 1, 0, 0)$ & $( 0, 0, 0, 0, 1, 1, 0, 0, 0)$ & $(1, 1, 0, 0, 1, 1, 0, 1, 1)$ & $12$      & $348$   & $24$         \\ \hline
\end{tabular}
\end{minipage}}
\end{table}

 \begin{table}[h!]\label{Type1-twelve}
\caption{New Binary Type I  Codes of Length 72 }
\resizebox{0.55\textwidth}{!}{\begin{minipage}{\textwidth}
\centering
\begin{tabular}{ccccccccccccccccc}
\hline
& Generator Matrix & $r_{A_{1}}$ & $r_{A_{2}}$& $r_{A_{3}}$ & $r_{A_{4}}$ & $r_{A_{5}}$ & $r_{A_{6}}$  & $r_{A_{7}}$  & $r_{A_{8}}$ & $r_{A_{9}}$ & $r_{A_{10}}$ & $r_{A_{11}}$ &$r_{A_{12}}$ & $\gamma$ & $\beta$ &$|Aut(C_i)|$ \\ \hline
$C_{34}$& $\mathcal{G}_7^{1}$ & $(0, 0, 1)$ & $(0, 1, 0)$ & $( 1, 1, 0)$ & $( 0, 1, 0)$ & $( 0, 1, 1)$ & $( 1, 0, 1)$ & $( 1, 1, 0)$ & $( 1, 0, 1)$ & $( 1, 1, 0)$ & $( 0, 0, 0)$ & $(1, 1, 1)$ & $(1, 1, 1)$ & $12$  & $373$   & $12$    \\ \hline
$C_{35}$& $\mathcal{G}_7^{2}$ & $(1, 1, 1,)$ & $( 1, 1, 1)$ & $( 1, 0, 0)$ & $( 1, 1, 0)$ & $( 0, 1, 0)$ & $( 1, 1, 1)$ & $(1, 1, 0)$ & $( 0, 1, 0)$ & $( 0, 1, 0)$ & $( 0, 1, 1)$ & $( 0, 1, 0)$ & $( 1, 0, 0)$ & $0$ & $161$   & $24$ \\ \hline
$C_{36}$ & $\mathcal{G}_7^{2}$ & $(1, 1, 1)$ & $( 1, 1, 0)$ & $( 1, 0, 1)$ & $( 1, 1, 0)$ & $( 0, 0, 0)$ & $( 1, 1, 0)$ & $(1, 1, 0)$ & $( 1, 0, 0)$ & $( 1, 1, 1)$ & $( 1, 0, 0)$ & $( 0, 1, 1)$ & $( 0, 1, 0)$ & $12$ & $301$   & $12$ \\ \hline
$C_{37}$ & $\mathcal{G}_7^{3}$ & $(0, 0, 1)$ & $( 1, 1, 1)$ & $( 1, 1, 0)$ & $( 0, 1, 1)$ & $( 0, 1, 1)$ & $( 1, 1, 1)$ & $( 1, 1, 0)$ & $( 1, 0, 1)$ & $( 1, 1, 0)$ & $( 0, 0, 0)$ & $( 1, 0, 0)$ & $( 1, 0, 0)$ & $0$  & $289$   & $12$   \\ \hline
$C_{38}$ & $\mathcal{G}_7^{4}$ & $(0, 0, 1)$ & $( 0,  1, 1)$ & $( 1, 1, 0)$ & $( 0, 1, 1)$ & $( 1, 0, 0)$ & $( 0, 1, 0)$ & $( 0, 0, 0)$ & $( 0, 0, 0)$ & $( 1, 0, 1)$ & $( 0, 0, 1)$ & $( 0, 0, 1)$ & $( 0, 0, 0)$ & $12$  & $253$ & $12$   \\ \hline
$C_{39}$ & $\mathcal{G}_7^{5}$  & $(1, 1, 0)$ & $( 0, 1, 0)$ & $( 0, 0, 1)$ & $( 1, 1, 0)$ & $( 1, 0, 1)$ & $( 1, 1, 0 )$ & $( 1, 0, 0)$ & $( 1, 1, 0)$ & $( 1, 0, 0)$ & $( 1, 1, 1)$ & $( 0, 0, 0)$ & $( 0, 0, 0)$ & $0$ & $260$   & $12$     \\ \hline
\end{tabular}
\end{minipage}}
\end{table}

\begin{table}[h!]\label{Type2}
\caption{New Binary Type II  Codes of Length 72}
\resizebox{0.65\textwidth}{!}{\begin{minipage}{\textwidth}
\centering
\begin{tabular}{cccccc}
\hline
& Generator Matrix       & $r_A$ & $r_B$ & $\alpha$ & $|Aut(C_i)|$ \\ \hline
$C_{40}$  &$\mathcal{G}_2^{3}$& $(1, 1, 0, 0, 0, 0, 1, 1, 0, 1, 1, 1, 0, 0, 1, 1, 1, 1)$ & $( 0, 0, 1, 0, 0, 1, 1, 1, 1, 1, 0, 1, 0, 1, 0, 0, 0, 0)$       & $-2796$  & $72$         \\ \hline
$C_{41}$  &$\mathcal{G}_2^{4}$& $(1, 0, 0, 0, 1, 0, 0, 0, 1, 1, 1, 1, 1, 1, 1, 0, 1, 0)$ & $(1, 0, 0, 1, 0, 0, 1, 0, 0, 0, 1, 0, 1, 0, 1, 1, 1, 1)$       & $-2598$  & $36$         \\ \hline
$C_{42}$  &$\mathcal{G}_2^{5}$& $(0, 1, 1, 1, 1, 1, 0, 1, 1, 0, 1, 1, 1, 1, 1, 1, 1, 0)$ & $(1, 0, 0, 0, 0, 0, 0, 1, 0, 1, 1, 1, 1, 1, 1, 0, 1, 0)$       & $-2388$  & $72$         \\ \hline
$C_{43}$ &$\mathcal{G}_3^{1}$ & $(1, 1, 0, 0, 0, 0, 0, 1, 1, 1, 1, 0, 1, 1, 1, 0, 1, 1)$  & $( 1, 0, 1, 1, 1, 0, 0, 0, 1, 0, 1, 1, 0, 1, 1, 1, 1, 1 )$  & $-3948$     & $144$         \\ \hline
$C_{44}$ &$\mathcal{G}_3^{2}$ & $(0, 1, 0, 1, 1, 0, 1, 0, 1, 0, 1, 0, 1, 0, 1, 0, 1, 1)$  & $(0, 1, 0, 0, 0, 0, 0, 0,1, 1, 1, 1, 1, 0, 0, 1, 1, 1)$  & $-2448$     & $144$         \\ \hline
$C_{45}$ &$\mathcal{G}_3^{2}$ & $(1, 0, 1, 1, 1, 1, 1, 0, 0, 1, 1, 1, 0, 0, 0, 1, 1, 1)$  & $( 0, 1, 1, 0, 1, 1, 1, 1, 1, 0, 0, 1, 0, 1, 0, 0, 1, 1)$  & $-2352$     & $144$         \\ \hline
$C_{46}$& $\mathcal{G}_3^{3}$ & $(1, 0, 1, 0, 1, 0, 0, 0, 0, 0, 0, 0, 1, 1, 1, 1, 0, 1)$  & $(0, 1, 1, 0, 0, 0, 1, 1, 0, 1, 1, 0, 0, 0, 0, 1, 0, 0)$  & $-4050$     & $72$         \\ \hline
$C_{47}$& $\mathcal{G}_3^{3}$ & $(1, 0, 1, 0, 0, 1, 1, 0, 1, 1, 1, 0, 1, 1, 1, 1, 1, 0)$  & $(1, 1, 1, 1, 1, 1, 1, 0, 1, 1, 1, 1, 1, 1, 1, 0, 1, 0)$  & $-4104$     & $144$         \\ \hline
$C_{48}$& $\mathcal{G}_3^{3}$&  $(0, 0, 0, 1, 1, 1, 1, 0, 1, 0, 0, 0, 1, 0, 1, 0, 0, 0)$  & $(1, 1, 1, 1, 1, 1, 1, 1, 1, 0, 1, 0, 0, 1, 0, 0, 1, 0)$  & $-2310$     & $144$         \\ \hline
$C_{49}$& $\mathcal{G}_3^{3}$ & $(0, 1, 1, 0, 1, 1, 0, 1, 0, 0, 0, 1, 1, 1, 1, 0, 0, 1)$  & $(0, 1, 0, 1, 0, 1, 1, 0, 1, 1, 0, 0, 1, 0, 0, 1, 0, 1 )$  & $-2292$     & $432$         \\ \hline
$C_{50}$& $\mathcal{G}_3^{4}$ & $(1, 1, 1, 1, 1, 0, 1, 0, 1, 0, 0, 1, 1, 0, 1, 1, 1, 0)$  & $(1, 0, 0, 1, 1, 1, 1, 0, 0, 1, 0, 0, 0, 0, 0, 1, 0, 0 )$  & $-2760$     & $432$         \\ \hline
$C_{51}$& $\mathcal{G}_3^{4}$ & $(0, 0, 0, 1, 1, 1, 0, 1, 1, 1, 0, 0, 1, 0, 0, 1, 1, 1)$  & $(1, 0, 1, 1, 0, 0, 0, 0, 1, 0, 0, 1, 0, 1, 1, 1, 0, 1)$  & $-2622$     & $72$         \\ \hline
$C_{52}$& $\mathcal{G}_5^{1}$ & $(0, 1, 0, 0, 1, 0, 0, 0, 1, 0, 0, 0, 1, 1, 1, 1, 1, 0)$  & $(0, 1, 0, 0, 1, 1, 0, 0, 1, 0, 1, 1, 0, 1, 0, 0, 0, 0 )$  & $-4020$     & $36$         \\ \hline
$C_{53}$ &  $\mathcal{G}_5^{3}$ & $(1,0,0,1,0,1,1,1,1,0,0,0,0,1,0,0,1,1)$  & $(0,1,1,0,1,0,1,1,1,0,1,0,0,1,1,0,0,1)$  & $-2688$     & $72$         \\ \hline
$C_{54}$& $\mathcal{G}_8^{2}$ & $(0, 1, 1, 1, 0, 1, 0, 1, 0, 0, 1, 1, 1, 1, 0, 0, 1, 1)$  & $(1, 1, 0, 1, 1, 0, 1, 1, 1, 0, 1, 1, 1, 0, 1, 0, 0, 1)$  & $-3966$     & $36$         \\ \hline
$C_{55}$& $\mathcal{G}_8^{3}$ & $(1, 0, 0, 1, 1, 0, 1, 1, 1, 0, 0, 0, 0, 0, 1, 1, 0, 0)$  & $( 0, 1, 0, 0, 0, 0, 1, 1, 1, 0, 0, 0, 0, 1, 0, 1, 0, 1 )$  & $-4056$     & $72$         \\ \hline
$C_{56}$& $\mathcal{G}_8^{3}$ & $(1, 0, 1, 0, 0, 0, 1, 1, 1, 0, 0, 0, 1, 1, 0, 1, 0, 0)$  & $(1, 1, 0, 1, 0, 1, 0, 0, 0, 1, 1, 1, 1, 0, 1, 1, 0, 1 )$  & $-3858$     & $144$         \\ \hline
\end{tabular}
\end{minipage}}
\end{table}

\begin{table}[h!]\label{Type2ABCD}
\caption{New Binary Type II  Codes of Length 72}
\resizebox{0.65\textwidth}{!}{\begin{minipage}{\textwidth}
\centering
\begin{tabular}{cccccccc}
\hline
& Generator Matrix      & $r_A$ & $r_B$ & $r_C$     & $r_D$  & $\alpha$ & $|Aut(C_i)|$ \\ \hline
$C_{57}$  &$\mathcal{G}_4^{2}$& $(0, 1, 1, 0, 1, 1, 0, 1, 0)$& $(0, 0, 1, 1, 1, 1, 0, 0, 1)$  & $( 0, 1, 0, 1, 0, 1, 1, 0, 1)$ & $(1, 0, 0, 1, 0, 0, 1, 0, 1 )$  & $-4080$     & $144$         \\ \hline
$C_{58}$ &$\mathcal{G}_4^{3}$& $(0, 1, 1, 0, 1, 1, 1, 1, 1)$& $(0, 1, 1, 0, 0, 1, 0, 0, 0)$  & $( 1, 0, 0, 1, 0, 1, 1, 1, 0)$ & $( 0, 0, 0, 0, 0, 0, 0, 0, 0 )$  & $-2238$     & $144$         \\ \hline
\end{tabular}
\end{minipage}}
\end{table}

\newpage

\section{Conclusion}

In this work, we applied the VOA to the problem of finding new binary self-dual codes by using a number of generator matrices of the form $[I_{36} \ | \ \tau_3(v)]$. We proved that the VOA outperforms the GA for the considered problem in terms of the number of codes found for the same constructions.
Moreover, with the matrix construction from \cite{Dougherty3} and the VOA, we found new binary self-dual codes. In particular,
we were able to construct 39 Type I binary $[72,36,12]$ self-dual codes with new weight enumerators in $W_{72,1}$:

\begin{equation*}
\begin{array}{l}
(\gamma =0,\ \  \beta =\{69, 75, 99, 129, 161, 236, 260, 289, 327, 359, 426\}),  \\
(\gamma =6,\ \  \beta =\{252, 276, 281, 341, 353 \}), \\
(\gamma =12,\ \  \beta =\{253, 267, 269, 285, 301, 314, 321, 348, 357, 373 \}), \\
(\gamma =18,\  \beta =\{279, 324  \}), \\
(\gamma =30,\ \  \beta =\{495 \}), \\
(\gamma =36,\ \beta =\{434, 456, 458, 468, 470, 483, 504, 506, 552, 633\}) \\
\end{array}%
\end{equation*}
and 19  Type II binary $[72,36,12]$ self-dual codes with new weight enumerators:
\begin{equation*}
\begin{array}{l}
(\alpha =\{-2238, -2292, -2310, -2352, -2388, -2448, -2598, -2622, -2688,  -2760, \\  -2796,  -3858, -3948, -3966, -4020, -4050, -4056, -4080, -4104   \}). \\

\end{array}%
\end{equation*}

\end{document}